\documentstyle[12pt,titlepage]{article}

\RequirePackage{amssymb}        %  extended mathematical character set

\title{Forms and Rotational States\\of the\\Nuclei of Ecliptic Comets}
\author{Colin~D.~B.~Snodgrass}
\date{September 2006}

\begin{document}

\maketitle

\section*{Abstract}

In this thesis I present measurements of the physical properties of the nuclei of Jupiter Family comets (JFCs), based on time-series observations. These data were collected in four observing runs; two using the 3.6m NTT in Chile, and two using the 2.5m INT on the island of La Palma. From the time-series photometry rotation rates and elongations were measured, and from these constraints were placed on the bulk density and porosity of nuclei. Multi-filter imaging was performed to enable measurement of their surface colours. In addition, a large amount of `snap-shot' imaging was performed during these observing runs, and taken with the time-series data is used to measure nuclei sizes. 

These results are compared with other data from the literature to study the general properties of JFC nuclei. A size distribution is measured which is consistent with that predicted for a population of collisional fragments, while the distribution in rotation rates is found to be flat and non-collisional. The low minimum densities measured for all comets imply that the true bulk density of nuclei is low, and the porosity is high. These properties are shown to have similar values in the Kuiper Belt Object (KBO) population, which is the supposed parent population for JFCs. The surface colours of JFCs are shown to match the blue end of the KBO distribution, and can be derived from the observed KBO distribution under the assumption of a de-reddening function that preferentially depletes the reddest surfaces.

\vspace{2cm}A pdf of the full thesis (3 MB) can be downloaded from: \\
http://homepage.mac.com/colinsnodgrass/FileSharing2.html

\end{document}